\documentclass[letter]{aa}     
\usepackage{graphicx}
\usepackage{txfonts}
\usepackage{lipsum}
\usepackage{subcaption}         
\usepackage{lscape}             
\usepackage{placeins}            
\usepackage[breaklinks=true]{hyperref} 
                                
\begin{document}

   \title{Assessing interstellar comet 3I/ATLAS with the 10.4 m Gran Telescopio Canarias and the 
          Two-meter Twin Telescope\thanks{Based on observations made with the Gran Telescopio 
          Canarias (GTC) telescope (program ID 62-GTC55/24B), in the Spanish Observatorio del 
          Roque de los Muchachos of the Instituto de Astrof\'{\i}sica de Canarias (IAC), and 
          the Two-meter Twin Telescope (TTT, PEI project PLANETIX25), in the Spanish Observatorio 
          del Teide of the IAC.}}

   \author{R. de la Fuente Marcos\inst{1}
        \and M. R. Alarcon\inst{2,3}
        \and J. Licandro\inst{2,3}
        \and M Serra-Ricart\inst{4,2,3}
        \and J. de León\inst{2,3}
        \and C. de la Fuente Marcos\inst{5}
        \and G. Lombardi\inst{6,2}
        \and A. Tejero\inst{6}
        \and A. Cabrera-Lavers\inst{6,2,3}
        \and S. Guerra Arencibia\inst{2,3}
        \and I. Ruiz Cejudo\inst{2,3}
        }
   \authorrunning{R. de la Fuente Marcos et al.}
   \titlerunning{Characterizing 3I/ATLAS with GTC and TTT} 
   \offprints{R. de la Fuente Marcos, \email{rauldelafuentemarcos@ucm.es}}
   \institute{
        AEGORA Research Group, Facultad de Ciencias Matem\'aticas, 
        Universidad Complutense de Madrid, Ciudad Universitaria, E-28040 Madrid, Spain 
        \and Instituto de Astrofísica de Canarias (IAC), C/Vía Láctea s/n, 38205 La Laguna, Tenerife, Spain
        \and Departamento de Astrofísica, Universidad de La Laguna, Avda. Astrofísico Francisco Sánchez, 38206 La Laguna, Tenerife, Spain
        \and Light Bridges, SL. Observatorio Astronómico del Teide. 
        Carretera del Observatorio del Teide, s/n, Güímar, Santa Cruz de Tenerife, Spain
        \and Universidad Complutense de Madrid, Ciudad Universitaria, E-28040 Madrid, Spain
        \and GRANTECAN, Cuesta de San José s/n, 38712 Breña Baja, La Palma, Spain\\}

   \date{Received 16 July 2025 / Accepted 24 July 2025}
 
  \abstract
   {Theories of the formation and evolution of small bodies in planetary 
    systems predict that they may escape into interstellar space at any 
    time, leaked from extrasolar Oort clouds or ejected following close 
    encounters with local planets or their host stars. After just two such interlopers -- 1I/2017~U1 (`Oumuamua) and 
    2I/Borisov -- had been characterized, more questions had been raised than answered. Assessing the 
    recently discovered interstellar comet 3I/ATLAS will only broaden our 
    understanding of this complex topic.}
   {Here, we investigate the spectral, cometary, and rotational properties 
    of 3I/ATLAS as well as its dynamical context.} 
   {We identified the spectral type of 3I/ATLAS from the visible reflectance 
    spectrum and used photometric observations to derive its level of 
    activity and rotational properties. Observational data were obtained with 
    the OSIRIS camera spectrograph at the 10.4~m Gran Telescopio Canarias and 
    the Two-meter Twin Telescope. We used $N$-body simulations and statistical 
    analyses of Gaia~DR3 data to investigate the origin of 3I/ATLAS and its 
    Galactic background.}
   {Interstellar comet 3I/ATLAS has a visible spectrum slightly redder than 
    the ones of D-type asteroids, 1I/'Oumuamua and 2I/Borisov, with a spectral 
    slope of $S'$=18.3$\pm$0.9~\%/1000~{\AA} in the 4000--9000~{\AA} range, which 
    is similar to the ones of trans-Neptunian objects and Centaurs. It has a conspicuous coma and its 
    rotation period is 16.79$\pm$0.23~h. The heliocentric components of its 
    Galactic velocity are $(U, V, W)=(-51.233{\pm}0.006,-19.456{\pm}0.004, 
    +18.930_{-0.006}^{+0.005})$~km~s$^{-1}$ with a radiant in Sagittarius of 
    $(\alpha,~\delta)=(295\fdg043_{-0\fdg004}^{+0\fdg003}, 
    -19\fdg0704_{-0\fdg0005}^{+0\fdg0006}$). The analysis of a sample of kinematic 
    analogs of 3I/ATLAS extracted from Gaia~DR3 suggests that its parent system 
    is part of the Galactic thin disk and includes a solar-like star with a slightly 
    subsolar metallicity.}
   {The results from the physical characterization of 3I/ATLAS further support
    the idea that extrasolar debris is not too different from the debris found in 
    the Solar System and that it is the result of similar formation processes.}


   \maketitle

   \nolinenumbers

\section{Introduction}
\label{sec:intro}
Interstellar objects are like cosmic messages in a bottle that convey information about the planetary systems where they formed 
for anyone to analyze, connecting distant worlds meant to never interact. Any debris in the form of asteroids and comets that 
escapes a mature planetary system will move with values of its heliocentric Galactic velocity components similar to the ones of 
its parent star. In the absence of significant scattering due to close encounters with other stars, this kinematic signature 
will remain relatively unaltered for some time. At least two comets -- C/1980~E1 (Bowell) and C/2024~L5 (ATLAS) -- have 
been recently ejected from the Solar System after experiencing planetary encounters at close range \citep{2024A&A...690A.395D}. 
In contrast, we had two visitors from beyond the Solar System -- 1I/2017~U1 (`Oumuamua) and 2I/Borisov -- and they left more 
questions raised than answered (see, e.g., \citealt{2017MPEC....V...17W,2017Natur.552..378M,2018Natur.559..223M,
2019RNAAS...3..131D,2020MNRAS.495.2053D,2025arXiv250705051D}). The study of more of these visitors from beyond the Solar System 
will only broaden our understanding of this subject.

Object A11pl3Z was initially reported on July 1, 2025, by the Asteroid Terrestrial-impact Last Alert System (ATLAS,  
\citealt{2018PASP..130f4505T}) observing from Rio Hurtado in Chile. It was announced as the third interstellar object on 
July~2 \citep{2025MPEC....N...12D} and designated 3I/ATLAS = C/2025~N1 (ATLAS). Follow-up and precovery observations 
\citep{2025MPEC....N...22Y,2025MPEC....N...35Y,2025MPEC....N...44V,2025MPEC....N...51C,2025MPEC....N...89B,2025MPEC....N..102C,
2025MPEC....O....5Y} revealed a highly eccentric, hyperbolic
orbit (eccentricity $e$$>$6). Here, we investigate observationally the spectral properties of 3I/ATLAS in the visible region 
using low-resolution spectroscopy, its cometary activity using high-S/N (signal-to-noise ratio) images, and its rotational state from photometric 
observations. Our goal is to place 3I/ATLAS within the context of both known interstellar objects and Solar System comets. In 
addition, we use its orbital solution to numerically explore its dynamical evolution. This paper is organized as follows. 
In Sect.~\ref{sec:results}, we analyze the observed properties of 3I/ATLAS, comparing them with the ones of known interstellar objects 
and Solar System comets. Our results are discussed in Sect.~\ref{sec:discussion} and our conclusions are summarized in Sect.~\ref{sec:conclusions}. Observations, data reduction, spectral extraction, and other technical details and input data are 
included as appendices.

\section{Results}
\label{sec:results}
Here, we use image enhancement algorithms applied to high-S/N images, reflectance spectroscopy, and photometric observations to
investigate the physical properties and rotational state of 3I/ATLAS. The dynamical status of this comet was studied using 
$N$-body simulations and its Galactic context was explored by applying data-mining techniques. Observational details are included in 
Appendices~\ref{sec:obs}, \ref{sec:appen_deep}, \ref{flux_rate}, and \ref{sec:appen_rot}. Appendix~\ref{ORBdata} provides information
on the numerical integrations.

\subsection{Cometary activity}
\label{sec:acti}
\begin{figure}[ht!]
\centering
\includegraphics[width=\linewidth]{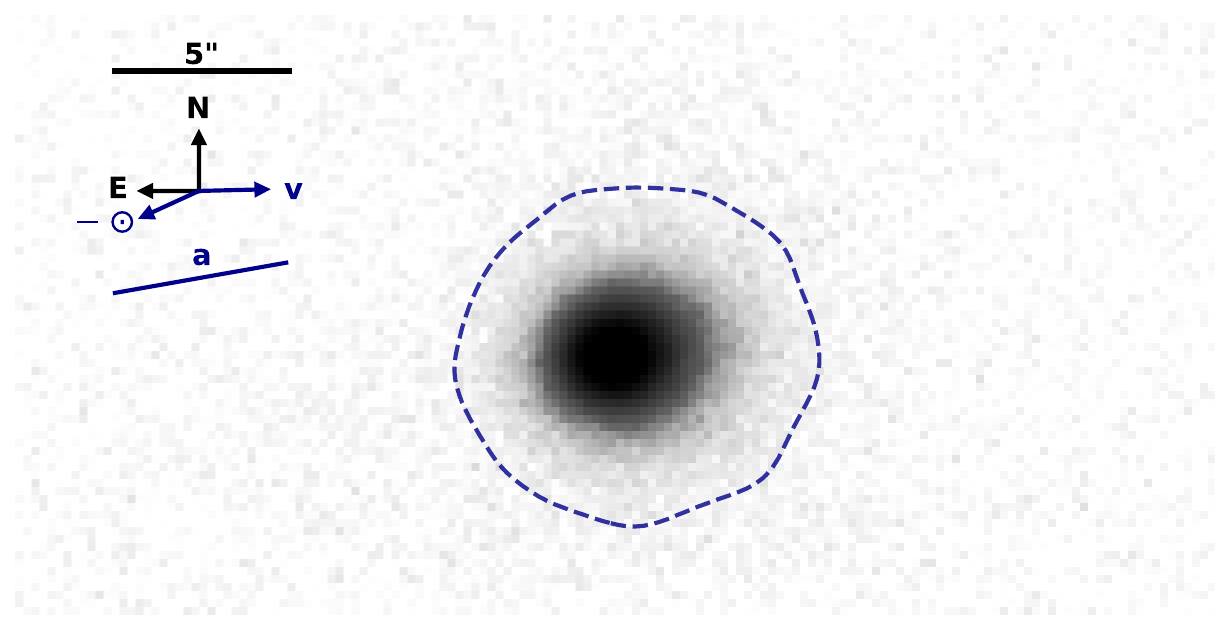}
\caption{Stacked $g$–band image created by co-adding 227 exposures (3.15 h) obtained on July 2, 2025. The overlaid contour delineates the coma boundary at the $2\sigma$ level above the sky background, corresponding to a surface brightness of $\mu_g = 25.4$ mag arcsec$^{-2}$ in the AB system. Arrows indicate the projected velocity vector of the comet ($v$) and the antisolar direction ($-\odot$). The direction of the semimajor axis ($a$) of the ellipse fit to the inner coma contours is also indicated.
\label{fig:coadd}}
\end{figure}

We generated a deep, stacked image, shown in Fig.~\ref{fig:coadd}, by co-adding 227 exposures (a total of 3.15 h) from the TTT telescope obtained on the night of 2 July. An overlaid contour delineates the coma boundary at the $2\sigma$ level above the background sky. At this threshold, the surface brightness of the edge is $\mu_g =25.4$ mag arcsec$^{-2}$ in the AB magnitude system. Details of the reduction and photometric calibration are given in Appendix~\ref{sec:appen_deep}.
The comet shows a well‐defined coma, extending over a projected area of $10.5'' \times 9.9''$. The major axis of this ellipse-like region is oriented at a phase angle of $279.5^\circ \pm 1.4^\circ$. At the time of observation, the geocentric distance was 3.44 au, so the physical dimensions were approximately 26 400~km $\times$ 24 700~km. 

Using a circular aperture of radius $4''$ ($\approx$10 000~km) from the centroid, we measured the integrated flux and converted it into the dust‑production proxy $A f\rho$ following the formulation of \citet{1984AJ.....89..579A}. After correcting this quantity for the $\alpha=2.53^{\circ}$ phase angle with the Halley–Marcus composite dust phase function \citep{2011AJ....141..177S}, we obtained a phase‑normalized $g$‑band value of $A(0)f\rho = 209 \pm 4$~cm.  This is slightly larger than the $187.7 \pm 4.1$~cm reported by \citet{10.1093/mnrasl/slaf078} for a comparable aperture.

\subsection{Spectral properties}
\label{sec:dust}
We studied the presence of typical cometary emission lines and bands in cometary comas (see e.g., \citealt{2012Icar..218..144C}). To do this, we proceeded as in \citet{2020MNRAS.495.2053D}, obtaining and analyzing the continuum-subtracted spectrum of 3I/ATLAS (see 
Fig.~\ref{flux_spectrum}). We first noticed that no cometary gas emission was detected, as was also reported by \cite{2025arXiv250707312A} and \cite{2025arXiv250705226O}; thus, the observed coma is mainly made of dust. Then we derived an upper limit for the most prominent gas emission band of comets in this spectral range, the CN (0--0) at 3880~{\AA}, and obtained $Q({\rm CN})_{\text{lim}}
=5.6{\times}10^{24}$~mol~s$^{-1}$ using the method described in Appendix~\ref{flux_rate}. Our results are consistent with the $Q({\rm CN})_{\text{lim}}$ value reported in \cite{2025arXiv250707312A}. Ours is larger because the $\sigma$ of their spectrum in the CN band spectral range is better as a result of their much higher spectral resolution. Measured CN production rates at large heliocentric distances are unusual \citep{1995Icar..118..223A} so 3I/ATLAS looks like a normal comet.

\begin{figure}[ht!]
\centering
\includegraphics[width=\linewidth]{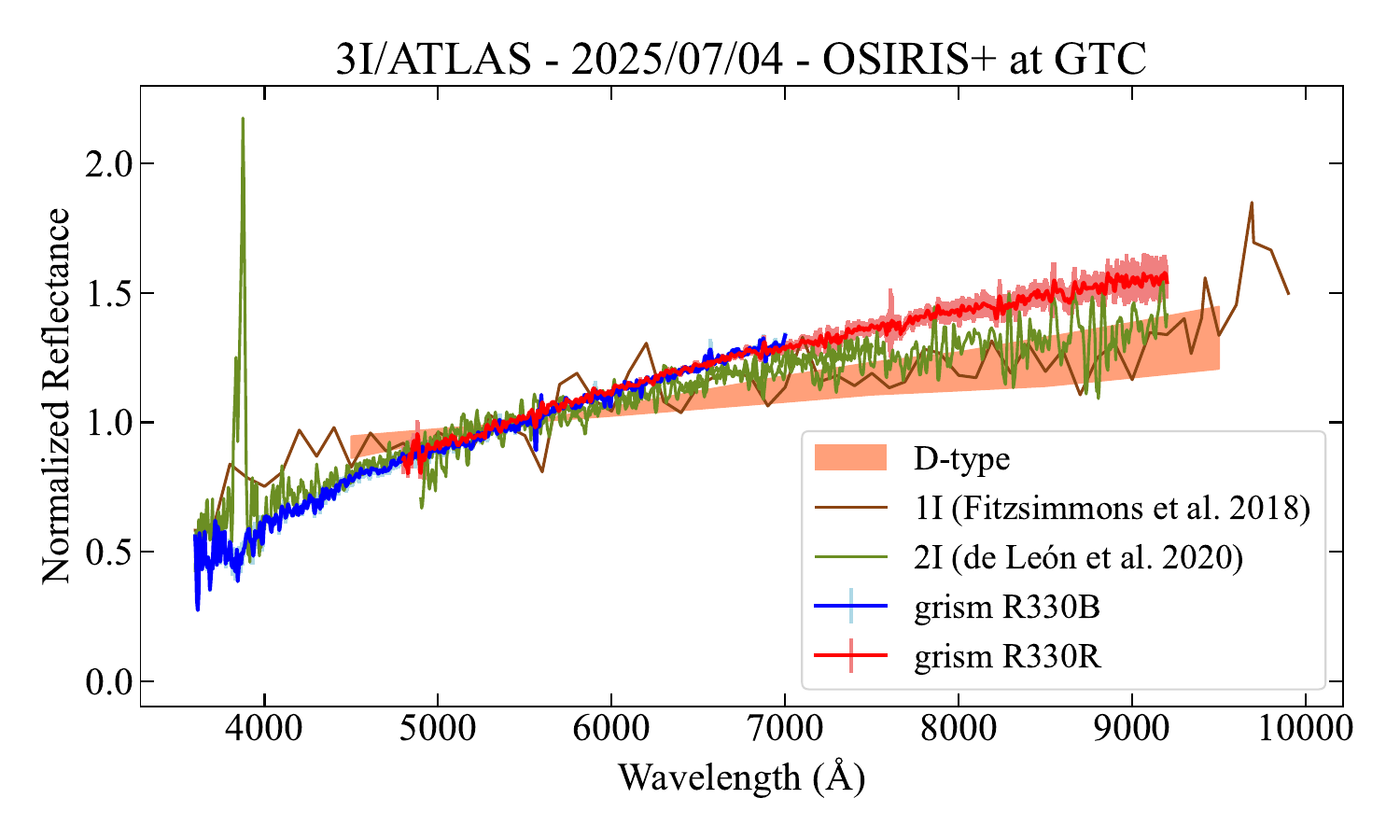}
\caption{Near-UV to visible reflectance spectrum of 3I/ATLAS obtained with OSIRIS at GTC and normalized to unity at 5500~{\AA}. Normalized reflectance of 1I/'Oumuamua (in brown) from \cite{2018NatAs...2..133F} and 2I/Borisov (in green) from \cite{2020MNRAS.495.2053D} are also shown for comparison, together with the D-type spectral taxon from \cite{2009Icar..202..160D}.
\label{fig:reflec}}
\end{figure}

On the other hand, the normalized reflectance spectrum of interstellar comet 3I/ATLAS, shown in Fig. \ref{fig:reflec}, is slightly redder than D-types, following \cite{2009Icar..202..160D}, and similar to the red spectra of the trans-Neptunian objects and Centaurs. For comparison purposes, we also show in the plot the normalized spectra of 1I/'Oumuamua \citep{2018NatAs...2..133F} and 2I/Borisov \citep{2020MNRAS.495.2053D}. We computed the reflectivity gradient, $S'$, by using a simple linear fit to the reflectance spectrum of 3I/ATLAS in the wavelength range 3800-9200 \AA, obtaining a value of $S$'=(18.3$\pm$0.9)\%/1000~{\AA}. The spectral slope error corresponds to the standard deviation of the mean of a total of 100 iterations, randomly removing 10 points during each fit. We are not considering here the systematic error of the observations associated with the use of two different solar analogs (see Appendix \ref{sec:spec}). Using the same wavelength intervals and normalization wavelength (6000~{\AA}) as in \cite{2025arXiv250705226O}, we obtain: (18.5$\pm$0.5)\%/1000~{\AA} in the 5000-7000~{\AA} range, (11.6$\pm$0.4)\%/1000~{\AA} in the 7000-9000~{\AA} range, and (14.6$\pm$0.2)\%/1000~{\AA} in the 5000-9000~{\AA} range. 

\subsection{Rotational state}
\label{sec:rot}
\begin{figure}[ht!]
\centering
\includegraphics[width=\linewidth]{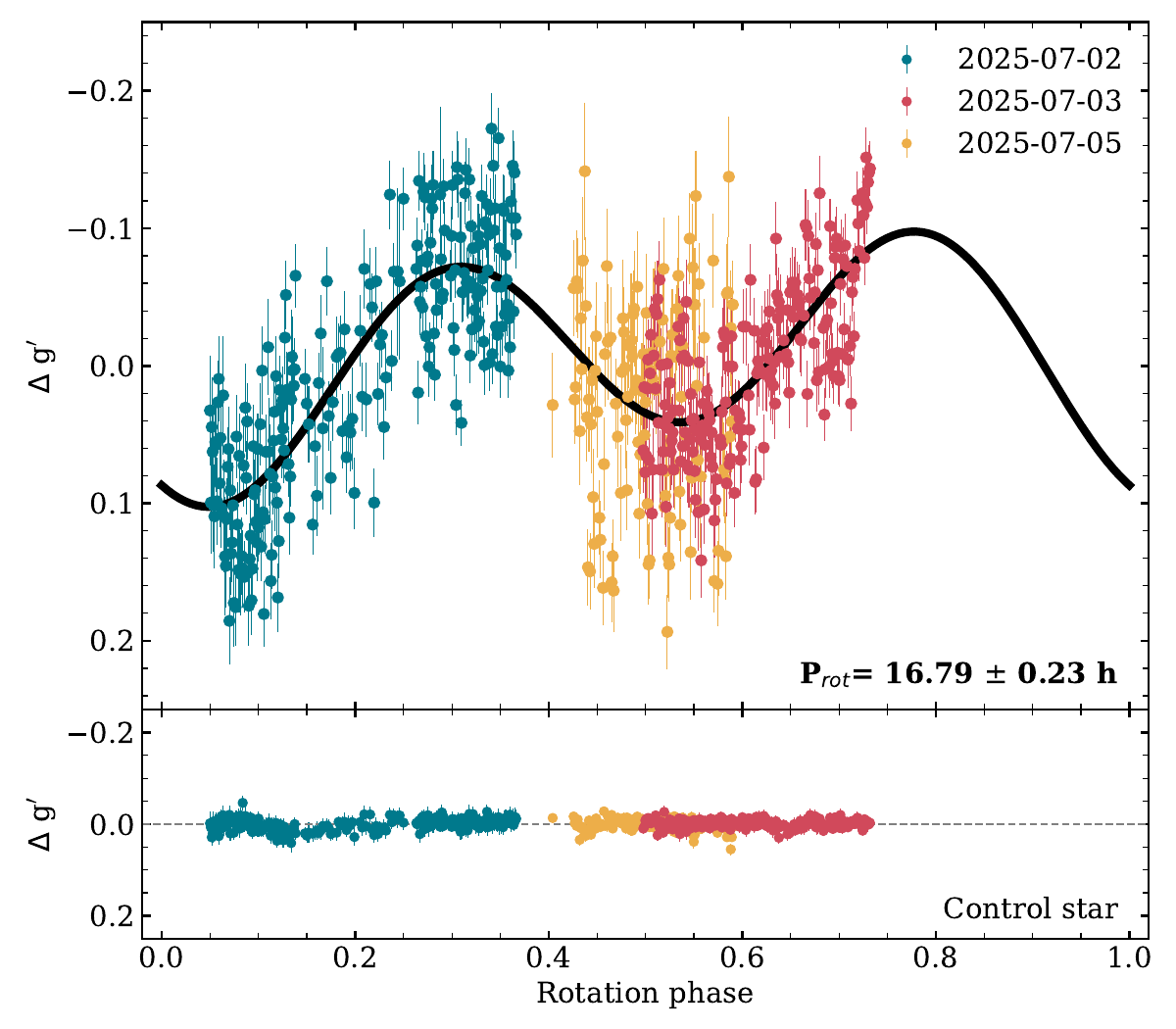}
\caption{Folded light curve of 3I/ATLAS. The top panel shows the magnitude variation relative to the nightly median, which is, in the g band, 18.76, 18.79, and 18.49, respectively. The black line corresponds to a second-order sinusoidal curve fit with a period obtained from the PDM method. The bottom panel displays the data for a non-variable control star, also normalized by its median magnitude, demonstrating the lack of significant variation in the photometric calibration.
\label{fig:lightcurve}}
\end{figure}

Time-series photometry is presented in Fig.~\ref{fig:lightcurve}, which includes observations from three nights (Table~\ref{tab:obs}). Data with excessively high noise due to poor seeing were removed from the plot, but were taken into account during the normalization of each light curve by subtracting the nightly median magnitude, which allowed them to be combined. The light curve shows a low amplitude of 0.2 magnitudes, which is typical for an object with cometary activity where the variability in light reflection from the nucleus's rotation is attenuated by the presence of a surrounding homogeneous coma \citep{2000AJ....119.3133L}. This is also consistent with observations in other preliminary works on 3I/ATLAS, such as \cite{Seligman2025}, in which, although they did not find a periodicity, they indicate that the brightness variations are small.
A clear trend is observed in the TTT data from each night of observation. We estimate the rotation period of 3I/ATLAS to be $P_{\text{rot}} = 16.79 \pm 0.23$ h. Further details are provided in Appendix~\ref{sec:appen_rot}.

\subsection{Dynamics}
\label{sec:dyna}
    The kinematic properties of interstellar interlopers are expected to preserve, for the most part, the kinematic signatures of their
    host stars. Analyzing the pre-encounter trajectory of 3I/ATLAS might shed some light on its true origin. We performed integrations
    backward in time of 1000 control orbits of 3I/ATLAS generated by the Monte Carlo using the covariance matrix (MCCM) method described 
    by \citet{2015MNRAS.453.1288D} and modified here to work with hyperbolic orbits. A statistical analysis (median and 16th and 84th percentiles) of the results indicates that at 1.7819$\pm$0.0003~pc from the Sun and 3$\times$10$^{4}$~yr into the past, 
    3I/ATLAS was moving inward, at $-$57.980$_{-0.008}^{+0.009}$~km~s$^{-1}$ -- 1I/2017~U1 (`Oumuamua) had $-26$~km~s$^{-1}$, in comparison \citep{2017RNAAS...1...21M}, and C/2019~Q4 (Borisov) had $-32.2815_{-0.0003}^{+0.0004}$~km~s$^{-1}$ \citep{2020MNRAS.495.2053D} --
    and projected toward (radiant or antapex) $\alpha=19^{\rm h}~40^{\rm m}~10.2^{\rm s}$, $\delta=-19\degr~04\arcmin~13.4\arcsec$ 
    (295\fdg043$_{-0\fdg004}^{+0\fdg003}$, $-19\fdg0704_{-0\fdg0005}^{+0\fdg0006}$) in the constellation of Sagittarius with Galactic coordinates $l=20\fdg726$, $b=-19\fdg057$, and ecliptic coordinates $\lambda=293\fdg603$, $\beta=+02\fdg340$. The heliocentric 
    components of its Galactic velocity were $(U, V, W)=(-51.233{\pm}0.006, -19.456{\pm}0.004, +18.930_{-0.006}^{+0.005})$~km~s$^{-1}$. These values were computed in the manner described by \citet{2019A&A...627A.104D}. The components of the solar-motion-corrected heliocentric Cartesian velocities of 3I/ATLAS are (62.3$\pm$0.8, $-$7.2$\pm$0.5, 26.2$\pm$0.4)~km~s$^{-1}$. We searched the {\it Gaia}~third~data~release (DR3, \citealt{2016A&A...595A...1G,2023A&A...674A...1G}), looking for kinematic analogs of 3I/ATLAS toward the region of the radiant. Figure~\ref{UVW} shows 3I/ATLAS in its Galactic kinematic context.

\section{Discussion}
\label{sec:discussion}
Our spectroscopic results are comparable with the ones reported by \citet{2025arXiv250705226O} and \citet{2025arXiv250707312A}, although we obtained a slightly bluer spectral slope in the 7000-9000~{\AA} range. Interestingly, \cite{2025arXiv250711720B} also found a more neutral slope for their spectrum in the range 7000--10000~{\AA}, as they did in  \cite{2025arXiv250712234K} at longer wavelengths. \citet{Seligman2025} report a slope of $S'$$\sim$18\%/1000~{\AA} from their $g$',$r$',$i$',$z$' colors and (17.1$\pm$0.2)\%/1000~{\AA} from their spectrum, consistent with the spectral gradient provided in this work in the 3800-9200~{\AA} range. The spectral slope of 3I/ATLAS is redder than the nominal values adopted by \cite{2023ARA&A..61..197J} for 1I/'Oumuamua (15$\pm$5 \%/1000~\AA) and 2I/Borisov (12$\pm$1 \%/1000~\AA). 

Considering \citet{2004oee..symp..154N}, fig.~1.3, the components of the solar-motion-corrected heliocentric Cartesian velocities
of 3I/ATLAS place it in the thin disk as the value of $(U_{\rm LSR}^{2} + W_{\rm LSR}^{2})^{1/2}$ is 67.58~km~s$^{-1}$ and 
$V_{\rm LSR}$=$-$7.2~km~s$^{-1}$. The Toomre diagram indicates that the thin disk is inside $(U_{\rm LSR}^{2} + W_{\rm LSR}^{2})^{1/2}
<$~85~km~s$^{-1}$, with the thick disk in the interval (85, 180)~km~s$^{-1}$ and the halo outside 180~km~s$^{-1}$ (see fig.~3 in \citealt{2003A&A...410..527B}). In addition to this kinematic constraint, the analysis of a sample of kinematic analogs of 3I/ATLAS 
from Gaia~DR3 gives a value of [Fe/H]=$-$0.04$\pm$0.14 (see Appendix~\ref{GaiaDR3}), which is somewhat subsolar, but far from 
$-$1.2$<$[Fe/H]$<$$-$0.4 in the thick disk (see, e.g. \citealt{2004oee..symp..154N}). An origin in the thick disk is favored by \citet{2025arXiv250705318H}. \citet{2025arXiv250708111T} argue for an age range of 3--11~Gyr and a value of [Fe/H]$=-0.18_{-0.21}^{+0.23}$, both of which are compatible with thin disk membership. 

\section{Conclusions}
\label{sec:conclusions}
In this paper, we have presented observations of interstellar comet 3I/ATLAS obtained with GTC and TTT. The GTC observations were used to derive the
spectral class and visible slope of 3I/ATLAS; the TTT ones provided the rotational period and insights into its cometary activity. Direct $N$-body simulations and statistical analyses were carried out to investigate the context of its orbital evolution. Our conclusions can be summarized as follows.
\begin{enumerate}
   \item High-S/N images clearly show that 3I/ATLAS is active at 3.44~au , with a  well‐defined dust coma that extends over a projected area of $10.5'' \times 9.9''$ ($26 400$~km $\times$ $24 700$~km at the comet's geocentric distance).  
   \item 3I/ATLAS has a visible reflectance spectrum slightly redder than the spectra of 1I/'Oumuamua and 2I/Borisov, but consistent with that of Solar System comets, trans-Neptunian objects, and Centaurs. Its overall spectral gradient in the 3800-9200~{\AA} range, $S$'=(18.3$\pm$0.9)\%/1000~{\AA}, is in good agreement with the findings of other authors, although somewhat bluer in the 7000-5000~{\AA} range ($S$'=11.6$\pm$0.4)\%/1000~{\AA}), suggesting a more neutral slope at longer wavelengths. 
   \item No signature of gas emission is observed in the visible spectrum of 3I/ATLAS, which is normal in Solar System comets observed at $r > 4$ au. We obtained an upper limit for the CN production rate of $Q({\rm CN})_{\text{lim}} = 5.6{\times}10^{24}$~mol~s$^{-1}$.
   \item The 3I/ATLAS light curve in the $g'$ band exhibits a low-amplitude variation with a period of 16.79$\pm$0.23~h.
   \item The analysis of an extensive set of $N$-body simulations indicates that 3I/ATLAS came from interstellar space and will return to it after reaching its perihelion on October 29, 2025. Its kinematic signature is consistent with an origin in the Galactic thin disk. 
\end{enumerate}
Follow-up observations of this object will provide further information on the properties of its parent planetary system.

\begin{acknowledgements}
We thank the anonymous referee for a prompt, constructive, detailed, and actionable report. 
We also thank Luisa Lara for her useful comments on the manuscript.
RdlFM acknowledges funding from the ``ENIMUS'' Advanced Grant from the European Research Council (ERC) under the 
European Union's Horizon 2020 research and innovation programme (grant agreement ID 101097905). This work was
partially supported by the Spanish `Agencia Estatal de Investigaci\'on (Ministerio de Ciencia e Innovaci\'on)' 
under grant PID2020-116726RB-I00 /AEI/10.13039/501100011033. This article is based on observations made in the 
Two-meter Twin Telescope (TTT\footnote{\url{http://ttt.iac.es}}) sited at the Teide Observatory of the Instituto 
de Astrofísica de Canarias (IAC), that Light Bridges operates in Tenerife, Canary Islands (Spain). The observation time rights (DTO) used for this research were consumed in the PEI 'PLANETIX25'. This research used storage and 
computing capacity in ASTRO POC's EDGE computing center at Tenerife under the form of Indefeasible Computer Rights 
(ICR). The ICR were consumed in the PEI "PLANETIX25" with the collaboration of Hewlett Packard Enterprise and VAST 
DATA. This work has made use of data from the European Space Agency (ESA) mission {\it Gaia} 
(\url{https://www.cosmos.esa.int/gaia}), processed by the {\it Gaia} Data Processing and Analysis Consortium (DPAC,
\url{https://www.cosmos.esa.int/web/gaia/dpac/consortium}). Funding for the DPAC has been provided by national 
institutions, in particular the institutions participating in the {\it Gaia} Multilateral Agreement.
\end{acknowledgements}

\bibliographystyle{aa}
\bibliography{3IATLAS}

\begin{appendix}

\section{Observations and Data Reduction}
\label{sec:obs}

Observations of interstellar comet 3I/ATLAS were carried out soon after its discovery and confirmation of interstellar nature, using the telescope facilities located at the Canarian Observatories (OCAN), managed by the Instituto de Astrofísica de Canarias (Spain). Observational circumstances are described in the following sections and summarized in Table \ref{tab:obs}.

\begin{table*}
\caption{Observational circumstances of the spectra and light-curves presented of 3I/ATLAS in this paper.}
\label{tab:obs}
\centering
\begin{tabular}{l c c c c c c c c}
\hline\hline
Observing date & N. images & UT$_{\rm start}$ & UT$_{\rm end}$ & Telescope & Instrument & Filter & $T_{\rm exp}$ (s) & $r$ (au) \\
\hline
2025-07-04 & 3   & 00:24 & 00:40 & GTC & OSIRIS+      & R300R & 300 & 4.42\\
2025-07-04 & 2   & 00:41 & 00:52 & GTC & OSIRIS+      & R300B & 300 & 4.41\\
2025-07-02 & 326 & 21:27 & 02:46 & TTT & Andor Ikon-L & $g$'    & 50 & 4.44 \\
2025-07-03 & 232 & 21:46 & 01:42 & TTT & Andor Ikon-L & $g$'    & 45 & 4.41\\
2025-07-05 & 307 & 21:29 & 03:12 & TTT & Andor Ikon-L & $g$'    & 50 & 4.34 \\
\hline
\end{tabular}
\tablefoot{The table includes the date of the observations, the total number of images (N. images), the starting and end times (UT$_{\rm start}$ and UT$_{\rm end}$), telescope, instrument, grating/filter used, exposure time of individual spectra/images ($T_{\rm exp}$), and object's distance to the Sun ($r)$.}
\end{table*}

\subsection{Spectroscopy}
\label{sec:spec}
Spectroscopic observations of 3I/ATLAS in the near-UV and visible wavelengths (3500--9200~{\AA}) were obtained using OSIRIS+ camera-spectrograph \citep{2000SPIE.4008..623C,2010ASSP...14...15C} at the 10.4m Gran Telescopio Canarias (GTC), located at the El Roque de Los Muchachos Observatory (La Palma, Spain). The OSIRIS+ instrument was upgraded in January 2023 to have enhanced sensitivity in the blue, down to 3700-3800 \AA. It is equipped with 2k$\times$4k pixel detector that provides a 7.8$\times$7.8 arcmin$^2$. We used the R300B (3600--7500~{\AA}, dispersion of 4.96~{\AA}~pixel$^{-1}$ for a 0.6" slit) and R300R (4800--9200~{\AA}, dispersion of 7.74~{\AA}~pixel$^{-1}$ for a 0.6" slit) grisms and a 1.5" slit width oriented in the parallactic angle. The telescope tracking was at the comet's proper motion. Two spectrophotometric standard stars (Feige 110 and Ross 640) were observed to flux calibrate the spectra. In the same way, to obtain the relative reflectance spectrum of the comet, we also observed two solar analog stars --- SA110-361 and SA112-1333 --- from the Landolt catalog \citep{1992AJ....104..340L}.

Data reduction included bias and flat-field correction, sky background subtraction, and one-dimensional aperture extraction, corresponding to the pixel where the peak intensity of the cometary profile decayed to its 10\% value. Wavelength calibration was then applied to the one-dimensional spectra using Xe+Ne+HgAr lamps. This procedure was applied to the spectra of the comet and the stars (both spectrophotometric and solar analogs). We divided the individual spectra of the comet by the spectra of the solar analogs, and the resulting ratio were averaged to get the final reflectance spectra. To create the final composite reflectance spectrum shown in Fig. \ref{fig:reflec}, we put together the R300B and R300R data using the common wavelength interval (6000--7000~{\AA}), and normalized the result to unity at 5500~{\AA}. The light shaded areas in the figure correspond to the standard deviation of the average, which introduces an error in the calculation of the spectral gradient $S$' smaller than 1\%/1000~{\AA}.  

\subsection{Imaging}
\label{sec:phot}
Photometric follow-up was carried out on the nights of July 2, 3 and 5, 2025, using one of the two telescopes of the Two-meter Twin Telescope (TTT), located at the Teide Observatory, Canary Islands (Lat. 28$^{\circ}$17$'$56$''$ N, Long. 16$^{\circ}$30$'$35$''$W, Alt. 2361.9 m). This is a 2-m $f$/6 Ritchey-Chrétien telescope that is currently in its commissioning phase. An Andor iKon-L 936 2k$\times$2k camera is mounted at the Nasmyth 2 focus, equipped with a back-illuminated 13.5~$\mu$m pixel$^{-1}$ BEX2-DD CCD sensor, resulting in a field of view of 7.85$'$$\times$7.85$'$ and a plate scale of 0.23$''$~pixel$^{-1}$.
The object transited the Galactic plane ($b$$\sim$1\fdg6, $l$$\sim$11\degr, on July 2), a region with both very high stellar density and significant differential extinction. Given that this extinction strongly affects bluer bands, we decided to observe with a $g'$ filter to reduce source blending. On the nights of July 3 and 5, a dithering pattern was used to avoid the potential effects of gradients across the sensor caused by the presence of the Moon and high airmass on the photometric quality of the data. A brief description of the image processing methods is included in Appendices~\ref{sec:appen_deep} and~\ref{sec:appen_rot}.

\section{Deep imaging data reduction
\label{sec:appen_deep}}
Each raw frame was first corrected for a master bias (median‐stack of 21 bias frames), saturated pixels were masked, and the flat-field correction was performed using twilight flats. A comprehensive sky mask using aggresive Gnuastro’s NoiseChisel \citep{Akhlaghi2015} segmentation was applied, and the sky level was measured via a sigma‐clipped median of the unmasked pixels and subtracted, ensuring removal of background without oversubtracting the faint coma.

An initial WCS solution was obtained using Astrometry.net \citep{Lang2010} against Gaia DR3, refined through three SCAMP \citep{Bertin2006} iterations. Each image was then cropped to a 3'$\times$3' region centered on the comet, using coordinates provided by JPL Horizons. To avoid potential astrometric uncertainties, the center was re-calculated from the object's brightness centroid. Cutout images were reastrometrized and resampled onto a common grid with SWarp (Lanczos3 method, \citealp{Bertin2002}). 
Photometric calibration was performed with reference to the Pan-STARRS~DR1 survey \citep{Chambers2016}.  
We cross-matched the TTT field with Gaia~DR3 \citep{2023A&A...674A...1G} to isolate bona fide stars, selected point sources by their FWHM, and performed aperture photometry with radii that enclose the PSF wings on both image sets, only retaining unsaturated stars with high signal-to-noise. Fluxes from both data sets were referenced to Gaia BP/RP spectra to remove the small zero‑point offset between surveys, and a colour correction based on the $g-r$ index and the respective filter–transmission curves accounted for filter‑response mismatches. The final scale factor places the TTT images on the AB system with a global zero‑point of 22.5 mag. this factor was applied to every sky‑subtracted frame before stacking, thereby converting pixel values from ADU to physical flux units.

After calibration, we examined all 326 frames for residual contamination (e.g., faint field stars or tracking artifacts) and excluded 99 images with any potential contamination to the comet signal. The remaining 227 frames, totaling 3.15 hours of integration, were combined using an inverse variance–weighted mean, where each image’s weight was set by the reciprocal of its background variance. In this combination, Noisechisel was used for masking every source but the comet and a 3$\sigma$ clipping was applied to reject outliers.

   \section{Gas production rate computation\label{flux_rate}}
\begin{figure}
    \centering
     \includegraphics[width=\linewidth]{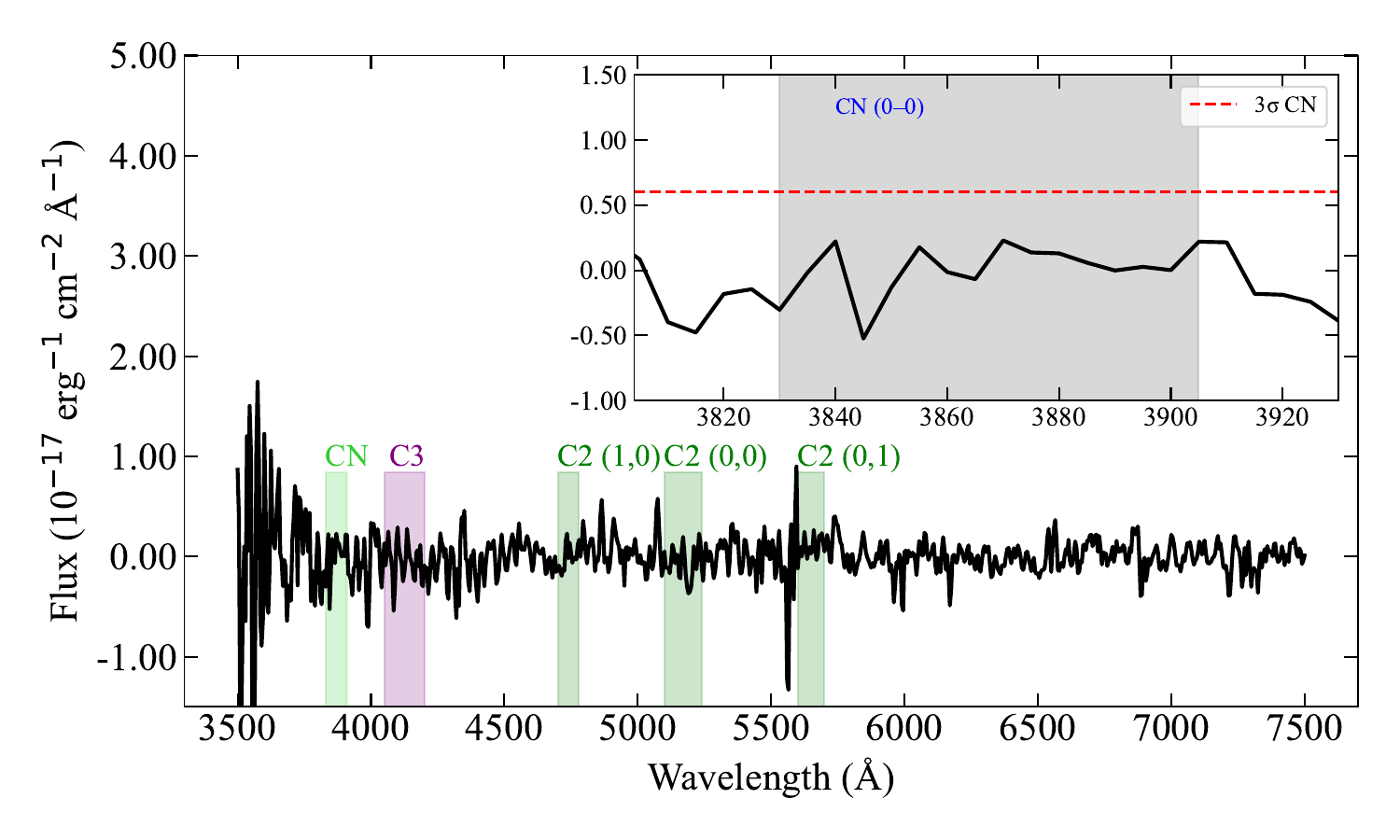}
     \caption{The flux spectrum of 3I/ATLAS obtained with OSIRIS at GTC, subtracted from the solar continuum. The most relevant regions were 
      CN, C$_3$ and c$_2$ bands in cometary spectra are shown. In the inset image, 
      a zoom of the CN most prominent cometary emission feature is shown. No significant signal is present in the data.}
      \label{flux_spectrum}
   \end{figure}

        To study the presence of typical cometary emission lines/bands of CN, C$_2$, C$_3$ or NH$_2$ usually detected in the visible range in active comets (see, e.g., \citealt{2012Icar..218..144C}), we first obtained the continuum-subtracted spectrum of 3I/ATLAS (see Fig.\ref{flux_spectrum}) using a spectrum of the Sun downloaded from the CALSPEC compilation \citep{2014PASP..126..711B} as we did in \cite{2020MNRAS.495.2053D}. Notice that the noisier part of the spectrum falls below 3800 \AA \ and thus it does not affect our computations. We first noticed that there is no evidence of any gas emission in our spectrum. As the CN (0--0) emission band at 3880 \AA\ use to be the most conspicuous one in comets, we computed the upper limit of the detectable CN flux within the S/N of our spectrum in the 3830--3905~{\AA} range (see Fig.\ref{flux_spectrum} inset plot) and derived the upper limit of the CN production rate (Q(CN)$_{lim}$) as follows: i) we first computed the $\sigma$ of the flux in the continuum-subtracted spectrum in the 3830--3905~{\AA} spectral range and obtained a value of $2.01\times10^{-18} \ \mathrm{erg\ s^{-1}\ cm^{-2}\ \mathring{A}^{-1}} $; ii) assuming a detection limit of 3-$\sigma$ as $F_{\text{lim}} = 3 \cdot \sigma \cdot \Delta\lambda = 4.53 \times 10^{-16} \ \mathrm{erg\ s^{-1}\ cm^{-2}}$; iii) we finally converted $F_{\text{lim}}$ into column density using the $g$ factor from \citet{2010AJ....140..973S}, scaled to both the heliocentric distance and the velocity of the comet. To compute the gas production rate, we assumed the Haser modeling with the outflow velocity $v_p$ scaled with $r_h$ ($v_p$=$0.86 r_h^{-4}$~km~s$^{-1}$), customary values for the daughter velocity $v_d$=1~km~s$^{-1}$ , and scale lengths given in \citet{1995Icar..118..223A}. We obtained $Q({\rm CN})_{\text{lim}} = 5.6\times 10^{24}$~mol~s$^{-1}$ .

\section{Period-finding method
\label{sec:appen_rot}}
All raw TTT frames were first corrected for bias and twilight flat‐field variations using standard procedures.  Point‐source photometry was then performed with a GPU‐accelerated custom pipeline that employs PSF‐matched convolutions for source detection and adaptive, S/N‐driven aperture photometry with PSF‐based aperture corrections.  
For each exposure, we matched field stars to the Pan-STARRS~DR1 catalog \citep{Chambers2016} to derive a photometric zero-point, which was used to calibrate the comet and control-star magnitudes in the resulting time series. This pipeline has been validated in previous works \citep{Popescu2025} and will be released as open-source software (Alarcon et al., in prep.).  

\begin{figure}[ht!]
\centering
\includegraphics[width=\linewidth]{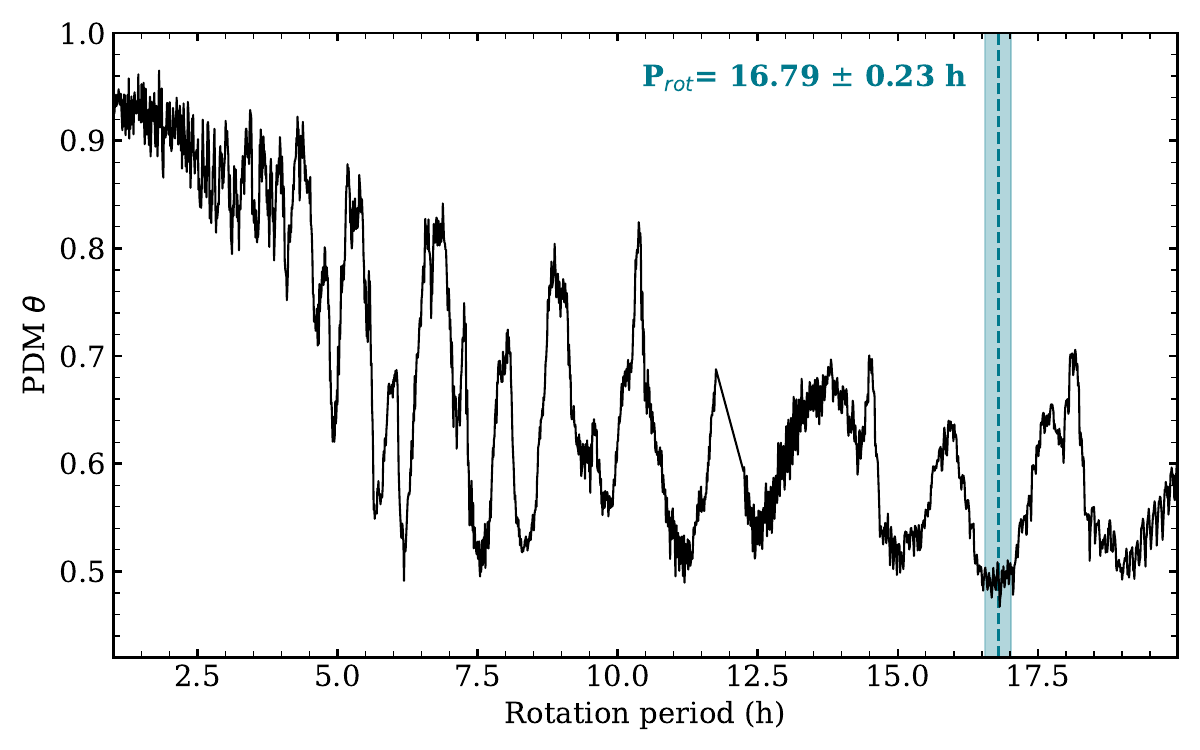}
\caption{Phase Dispersion Minimization (PDM) Analysis. The PDM theta statistic is plotted against the trial period, with the global minimum identifying the best-fit period of 16.79 $\pm$ 0.23 hours, The 1-sigma (FWHM/2.355) uncertainty region, marked by the blue shaded band, was derived from a gaussian fit to the minimum.
\label{fig:pdm}}
\end{figure}

To determine the rotation period of 3I/ATLAS from the TTT time‐series photometry, we began by computing a generalized Lomb–Scargle periodogram \citep{Press1989}. This method is well suited for unevenly sampled data and quantifies the significance of sinusoidal signals by fitting 2$^{\text{nd}}$-order sine curves at trial frequencies while properly normalizing for observational errors.  Applying it to the combined $g'$‐band light curve revealed a peak at $P_{\text{rot}} = 16.66$ h. To corroborate this result without assuming purely sinusoidal variability, we employed the phase dispersion minimization (PDM) technique, which searches for the period that minimizes the variance of data points within phase bins. Our PDM statistic exhibited a global minimum at a similar period identified by the Lomb–Scargle analysis.  We then fit a Gaussian function to the PDM curve in the vicinity of this minimum (see Fig.~\ref{fig:pdm}), yielding a refined rotation period of $P_{\text{rot}} = 16.79 \pm 0.23$ h. The quoted uncertainty corresponds to the standard deviation of the Gaussian fit and encompasses both measurement noise and sampling effects.  We adopt this value as our best estimate in the main text.

To verify that the observed periodic modulation originates from the comet’s nucleus and coma rather than observational artifacts, we investigated potential correlations between measured flux and atmospheric seeing, as well as any drifts in our photometric zero point.  In addition, we extracted light curves for several field stars of similar brightness; the bottom panel of Fig.~\ref{fig:lightcurve} presents a representative non‐variable control star, which shows scatter an order of magnitude smaller than the amplitude of the comet’s modulation. These checks confirm that our period determination is robust against changes in seeing and calibration, and that the periodic signal reflects the intrinsic rotation of 3I/ATLAS.

   \section{Orbit data and post-perihelion evolution\label{ORBdata}}
      The orbital solution in Table~\ref{elements} is currently based on 762 observations with an observational time span of 
      60~days, it is referred to epoch JD 2460868.5 TDB, which is the origin of time in the calculations. It was retrieved from Jet Propulsion Laboratory's (JPL) Small-Body Database 
      (SBDB)\footnote{\href{https://ssd.jpl.nasa.gov/tools/sbdb\_lookup.html\#/}{https://ssd.jpl.nasa.gov/tools/sbdb\_lookup.html\#/}} provided by the Solar System Dynamics Group (SSDG, \citealt{2011jsrs.conf...87G,
      2015IAUGA..2256293G}).\footnote{\href{https://ssd.jpl.nasa.gov/}{https://ssd.jpl.nasa.gov/}} Its current path is hyperbolic with a 
      statistical significance (barycentric) above 1718$\sigma$. No known processes, close encounters with known planets included, can 
      explain such a large eccentricity if we were to assume that 3I/ATLAS has an origin in the Solar System; an extrasolar origin 
      emerges as the single other alternative.

     \begin{table}
        \fontsize{8}{12pt}\selectfont
        \tabcolsep 0.15truecm
        \caption{\label{elements}Heliocentric and barycentric orbital elements and 1$\sigma$ uncertainties of comet 
                  C/2025N1~(ATLAS).
                }
        \centering
        \begin{tabular}{lccc}
           \hline\hline
            Orbital parameter                                 &   & Heliocentric         & Barycentric \\
           \hline
            Perihelion distance, $q$ (au)                     & = &    1.3570$\pm$0.0005 &    1.3623   \\
            Eccentricity, $e$                                 & = &    6.144$\pm$0.003   &    6.156    \\
            Inclination, $i$ (\degr)                          & = &  175.1135$\pm$0.0002 &  175.1275   \\
            Longitude of the ascending node, $\Omega$ (\degr) & = &  322.164$\pm$0.005   &  322.254    \\
            Argument of perihelion, $\omega$ (\degr)          & = &  128.007$\pm$0.004   &  128.086    \\
            Mean anomaly, $M$ (\degr)                         & = & -796.3$\pm$0.3       & -795.8      \\
           \hline
        \end{tabular}
         \tablefoot{The orbit solution has been computed by D. Farnocchia at epoch JD 2460868.5 that corresponds to 
                   00:00:00.000 TDB, Barycentric Dynamical Time, on 2025 July 12, J2000.0 ecliptic and equinox and 
                   it is based on 762 observations with an observational time-span of 60 days. 
                   Source: JPL's SSDG SBDB (solution date, 2025-Jul-21 13:23:29 PDT).
                  }
     \end{table}

    Answering the question of where 3I/ATLAS is headed next after its flyby with the Sun requires the analysis of a similar set of 
    direct $N$-body simulations (also with MCCM), but forward in time. At  1.7823$\pm$0.0003~pc from the Sun and 
    3$\times$10$^{4}$~yr into the future, this object will be receding from us at $57.994_{0.008}^{0.009}$~km~s$^{-1}$ toward (apex) 
    $\alpha=06^{\rm h}~20^{\rm m}~55.3^{\rm s}$, $\delta=+19\degr~48\arcmin~14.0\arcsec$ (95\fdg23$\pm$0\fdg06, 19\fdg8039$\pm$0.0005) 
    in the constellation of Gemini with Galactic coordinates $l=191\fdg904$, $b=02\fdg4825$, and ecliptic coordinates 
    $\lambda=94\fdg929$, $\beta=-03\fdg546$. The heliocentric components of its Galactic velocity will be $(U, V, W)=(-56.679\pm0.007, -12.018\pm0.008, +2.512_{0.013}^{0.014})$~km~s$^{-1}$.

   \section{\textbf{\itshape{Gaia}} DR3 sample\label{GaiaDR3}}  
      An origin in the Galactic thin disk is supported by the fact that several stars in {\it Gaia}~third~data~release (DR3) with 
      properties consistent with thin disk membership have kinematic signatures compatible with that of 3I/ATLAS. Gaia~DR3 provides
      estimates of $T_{\rm eff}$ and [Fe/H] for some of the relevant stars and the values are consistent with those of 
      Solar-like stars of Solar-like metallicity as shown in Fig.~\ref{thindisk}. One star (source 4091535895147524608) with 
      3I/ATLAS-like kinematics has $\alpha$ = 18:34:31.4, $\delta$=$-$21:39:30, $T_{\rm eff}$=4580.2~K, [Fe/H]=0.20, a heliocentric 
      distance of 2232.16~pc, and heliocentric Galactic velocity $(U, V, W) = (-51\pm4, -20.1\pm1.3, +18\pm2)$~km~s$^{-1}$.
      Unfortunately, the uncertainties in the values of the radial velocity of the relevant stars in {\it Gaia}~DR3 are significant
      and this leads to errors in the components of the velocity which are much larger than those computed for 3I/ATLAS.

   \begin{figure}
    \centering
     \includegraphics[width=\linewidth]{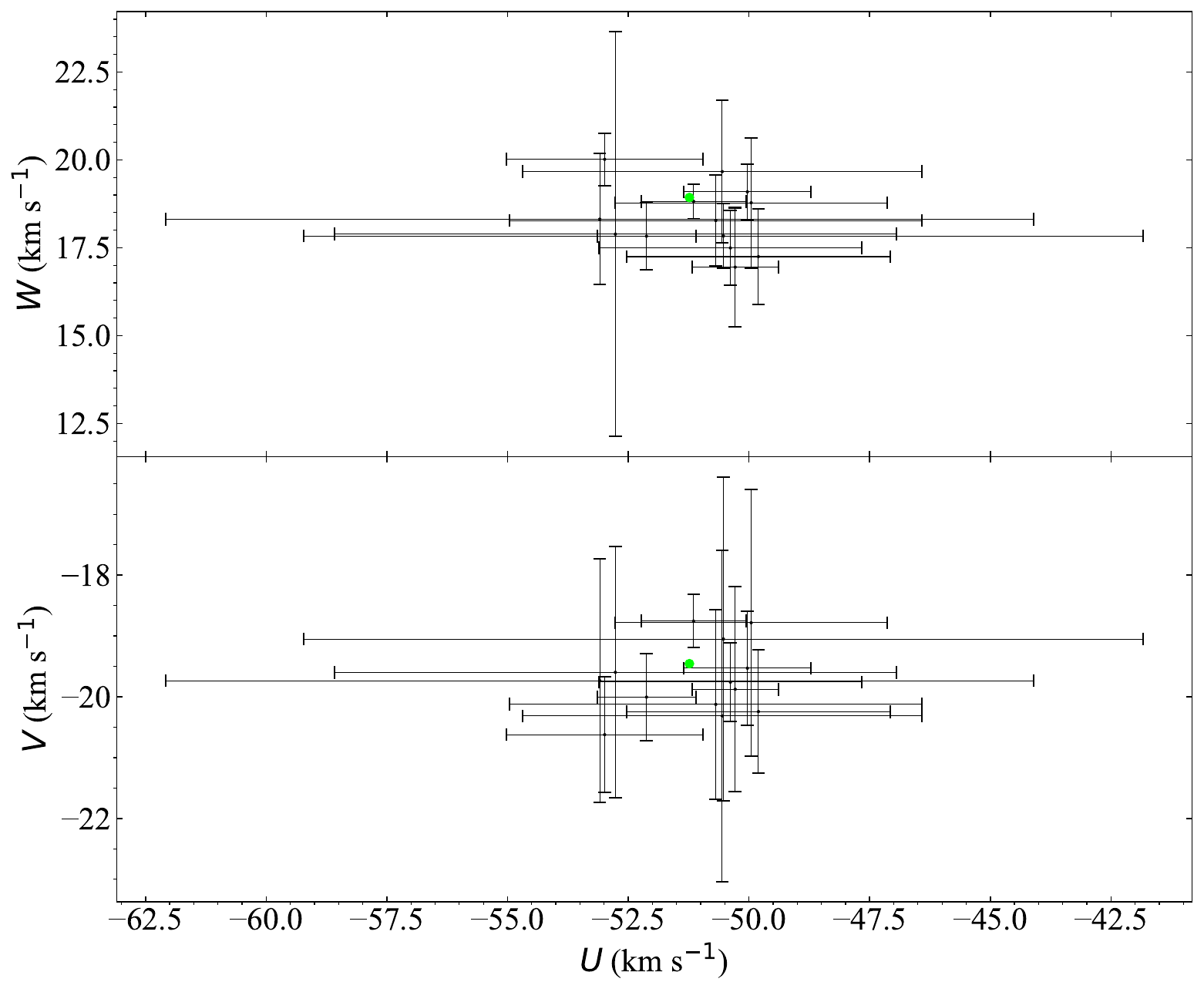}
     \caption{Galactic kinematic context of 3I/ATLAS. {\it Gaia}~DR3 stars with values of their heliocentric Galactic velocity 
              components close to those of 3I/ATLAS in black, 3I/ATLAS in green.}
     \label{UVW}
   \end{figure}
   \begin{figure}
    \centering
     \includegraphics[width=\linewidth]{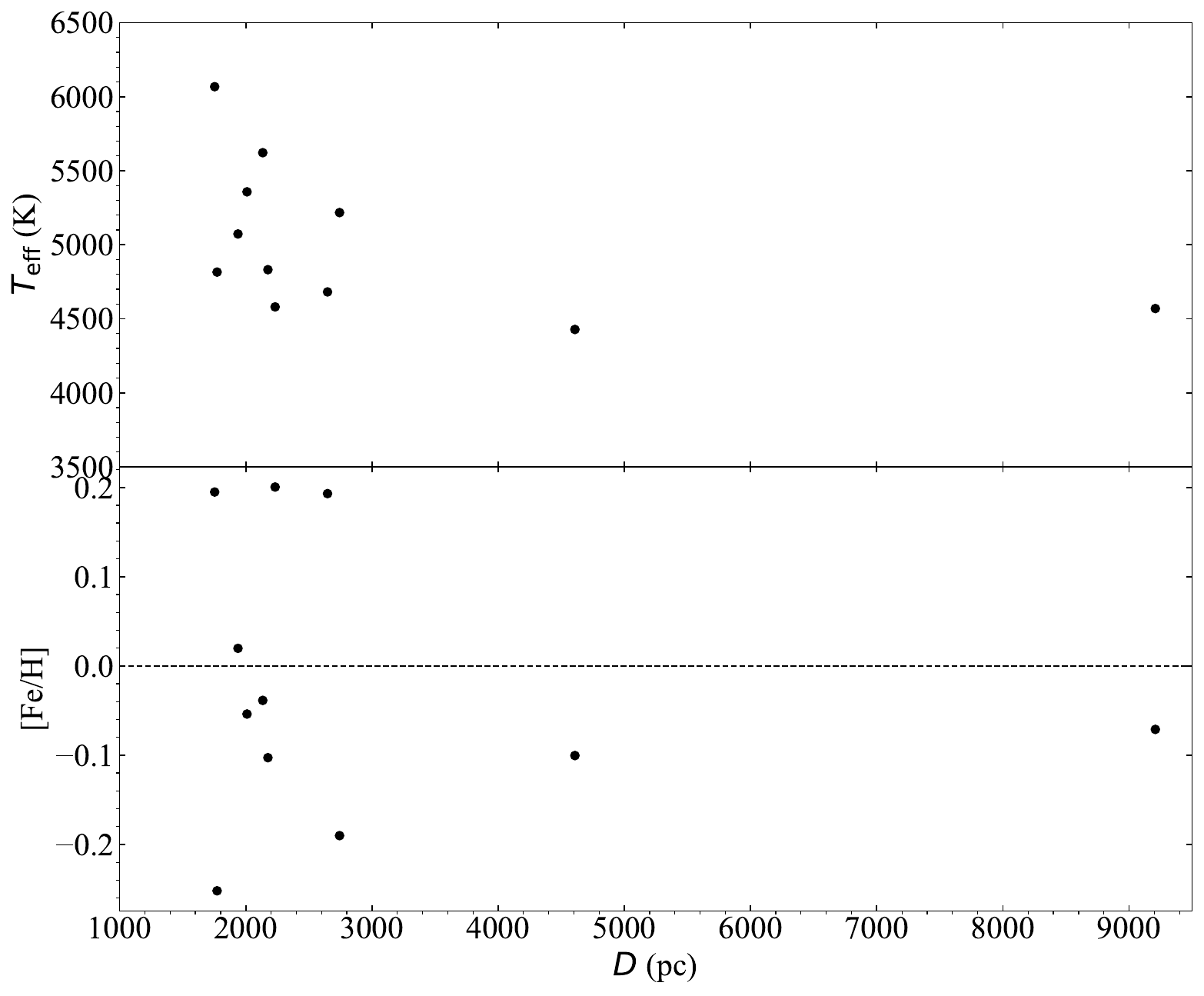}
     \caption{Kinematical analogs of 3I/ATLAS: Relevant properties. {\it Top panels:} {\it Gaia}~DR3 estimated values of effective
              temperature and heliocentric distance. {\it Bottom panel:} {\it Gaia}~DR3 estimated values of metallicity and
              heliocentric distance.}
     \label{thindisk}
   \end{figure}

\end{appendix}

\end{document}